\documentclass[10pt]{iopart}
\usepackage{epsfig}
\usepackage{graphicx}
\usepackage{dcolumn}
\usepackage{iopams}

\expandafter\let\csname equation*\endcsname\relax 
\expandafter\let\csname endequation*\endcsname\relax 

\usepackage{amsmath}
\usepackage[english]{babel}
\usepackage[utf8]{inputenc}
\usepackage{hyperref}
\usepackage{bm}

\newcommand{\op}{\mathrm{open}} 
\newcommand{\cl}{\mathrm{closed}}

\begin{document}

\title[Finite Range Effects for Three Bosons]{Finite Range Effects in Energies and Recombination Rates of Three Identical Bosons}

\author{P K S{\o}rensen, D V Fedorov, A S Jensen and N T Zinner}
\address{
	Department of Physics and Astronomy - Aarhus University, Ny Munkegade, bygn. 1520, DK-8000 \AA rhus C, Denmark
}

\date{\today}
 
\begin{abstract}
We investigate finite-range effects in systems with three 
identical bosons. We calculate recombination 
rates and bound state spectra using two different finite-range models 
that have been used recently to describe the physics of
cold atomic gases near Fesh\-bach resonances where the 
scattering length is large. 
The models are built on contact potentials 
which take into account finite range effects; 
one is a two-channel model and the other is an effective
range expansion model implemented through the boundary condition 
on the three-body wave function when two of the particles
are at the same point in space.
We compare the results with the results of the ubiquitous 
single-parameter zero-range model where only
the scattering length is taken into account. Both finite
range models predict variations of the well-known geometric 
scaling factor 22.7 that arises in Efimov physics. 
The threshold value at negative scattering length 
for creation 
of a bound trimer moves to higher or lower values depending 
on the sign of the effective range compared to the location 
of the threshold for the single-parameter zero-range model.
Large 
effective ranges, corresponding to narrow resonances, are 
needed for the reduction to have any noticeable effect.
\end{abstract}
\maketitle

\section{Introduction}\label{Introduction}
Low-energy quantum mechanical bound states present a number of surprising results 
when the state contains three or more particles. The most famous case is the Efimov effect that occurs when
a three-body system has two-body subsystems that have a two-body bound state with zero energy. In this 
situation one can mathematically prove that an infinite number of three-body states appear when 
the particles are bosons \cite{efi70}. More generally, the effect can also take place for two-component 
fermionic systems whenever the mass ratios are large enough \cite{Nielsen2001,jen04,braaten2006}. In recent 
years, it has become clear that few-body effects such as this can be studied in great detail in 
experiments with ultracold atomic gases via tunable Feshbach resonances \cite{ferlaino2010,chin2010}. Producing
a two-body bound state at zero energy is thus routinely done and 
signatures of three- and even four-body resonances have been observed \cite{kraemer2006,pollack2009,zaccanti2009,gross2009,williams2010,gross2010,lompe2010,japan2011,berninger2011,wild2012,knoop2012}

The few-body features that are studied in ultracold atom experiments are often identified through the rate at which atoms are lost from the experimental trapping potential. In fact, the densities and lifetimes of typical Bose-Einstein-Condensates (BEC) are limited by loss effects, primarily due to three-body recombination, a process where three particles interact and create a bound system of two particles (dimer) and the third particle carries away excess energy and momentum, generally resulting in a loss of all three particles from the confining trap. The loss rate is given by $\dot n=-\alpha n^3$ where $n$ is the particle density and the recombination coefficient is $\alpha=C(a)\hbar a^4/m$ with $C(a)$ a log-periodic function of the two-body scattering length $a$ with period 22.7 \cite{Nielsen2001}. A strongly interacting BEC can be created by use of Feshbach resonances where the scattering length can be tuned using an external magnetic field \cite{chin2010}.

The $a^4$ scaling of the recombination coefficient is valid when the scattering length is much larger than the range of the inter-atomic potential. Here we use different models to estimate the effect of the finite range of the interactions. All the models we use are based on contact interactions but, in contrast to the typical one-parameter implementation, we also incorporate a non-zero effective range. This is done by either modifying the Bethe-Peierls boundary condition using the effective range expansion or by using a two-channel model which not only has an inherent effective range but also qualitatively describes the physics of Fesh\-bach resonances.

The hyperspherical adiabatic approach is used to handle the three-particle problem. This method works equally well for the three 
scattering models we consider (zero-range contact interaction, effective range expansion and two-channel contact interaction). In order
to calculate the recombination rate, we use the WKB-method with hidden crossings \cite{Macek1999} that takes the three-body
system from the initial in-coming scattering channel into the atom plus dimer channel of lower energy via an excursion into the complex
plane. Since this approach depends critically on the presence of a bound dimer, it is only applicable to the positive $a$ side of a Feshbach resonance and we only consider recombination rates for $a>0$. Along the way, we discuss the hyperspherical potentials for the different models, and also the scaling properties of consecutive Efimov resonances when effective range corrections are included.

In order to also address the negative $a$ side of the resonance, we calculate the spectrum of bound low-energy trimers within 
different models and study their dependence on the effective range corrections. Recent experiments have shown that the 
lowest three-body Efimov state has a breakup threshold at the three-atom continuum on the $a<0$ side of the resonance that seems
to be uniquely determined by the background van der Waals parameter of the atomic system under study \cite{berninger2011}. In fact,
the threshold, $a^{(-)}$, is experimentally found to fulfil $a^{(-)}\sim -9.8r_\textrm{vdW}$, where $r_\textrm{vdW}$ is the two-body
van der Waals length scale \cite{berninger2011}. This finding has induced a number of theoretical studies that aim to explain the 
proportionality and calculate the constant factor from basic knowledge about the two-body atomic potentials 
\cite{naidon2011,chin2011,schmidt2012,wang2012,peder2012,naidon2012}. 
It can be most easily understood as consequence of the hard-core repulsion of the atom-atom interaction (typically
of Lennard-Jones type with a corresponding van der Waals tail). This fact induces a hard-core repulsion in the 
three-body potential around $r_\text{vdW}$ as shown numerically in Ref.~\cite{wang2012} and soon after by
analytical means in Ref.~\cite{peder2012}. Here we calculate the three-body spectrum for different scattering 
models and study the influence of effective range corrections on the threshold and the scaling properties
of the states. Finite range corrections beyond the scattering length approximation have been considered 
before within various approaches \cite{fj01b,kokkelmans02,fj02,jonsell2004,petrov2004,tfj08c,pjp09,tfj09,jpp10,wang11,zinner2011}.
However, to the best of our knowledge systematic studies using simple zero-range models have not been presented. Since
zero-range models are extremely convenient in both few- and many-body studies, it is important to gauge their
applicability which is one goal of the current study.

The paper is organized as follows: In Sec.~\ref{hyper} we first present the hyperspherical adiabatic method 
which constitutes our theoretical framework and discuss the differences in the solutions to the 
hyperspherical three-body potential within the different two-body interaction models that we employ. 
Section~\ref{recRate} discusses the recombination rate within the different models using the hidden crossing technique.
In section~\ref{boundTrimers} we discuss the spectrum of three-body bound states with the various models and 
map out the dependence on the effective range. Section~\ref{Conclusion} contains a short summary along with conclusions and outlook.

\section{Formalism}\label{hyper}
We consider a system of three identical bosons of 
mass $m$ using hyper\-spherical coordinates defined from 
the Cartesian coordinates $\bm r_i$,$\bm r_j$,$\bm r_k$, of particles $i,j,k$ as
\begin{align}
	\bm x_i&=\frac{\bm r_j-\bm r_k}{\sqrt2},\quad\bm y_i=\sqrt{\frac23}\left(\bm r_i-\frac{\bm r_j+\bm r_k}{2}\right)\;,
	\label{eq:1}\\
	\rho^2&=|\bm x_i|^2+|\bm y_i|^2,\qquad\tan\alpha_i=\frac{|\bm x_i|}{|\bm y_i|}\;,
	\label{eq:2}
\end{align}
where $\{i,j,k\}$ are cyclic permutations of $\{1,2,3\}$, $\rho$ is the hyperradius and $\alpha_i$ is a hyperangle \cite{Nielsen2001}. The directions of $\bm x_i$ and $\bm y_i$ comprise four additional hyperangles which, along with $\alpha_i$, are denoted collectively as $\Omega$. The hyperradius $\rho$ is independent of the choice of $\{i,j,k\}$.

The wave function is expanded on adiabatic basis states $\Phi_n(\rho,\Omega)$. 
In the hyperspherical adiabatic approximation only the first term in this 
expansion is kept, this has been proven to be a good approximation when 
dealing with Efimov three-body states (trimers) \cite{Nielsen2001}. The wave function is then
\begin{equation}
	\Psi(\rho,\Omega)=\rho^{-5/2}f_0(\rho)\Phi_0(\rho,\Omega),
	\label{eq:3}
\end{equation}
where $\Phi_0(\rho,\Omega)$ is  the solution to the hyperangular equation
\begin{equation}
	\left(\Lambda+\frac{2m\rho^2}{\hbar^2}V\right)\Phi_0(\rho,\Omega)=\lambda_0(\rho)\Phi_0(\rho,\Omega),
	\label{eq:4}
\end{equation}
where $\Lambda$ is the grand angular momentum operator in hyperradial coordinates (see \cite{Nielsen2001}) 
and $\lambda_0$ is the corresponding eigenvalue. The hyperradial function $f_0(\rho)$ is a solution to the hyperradial equation
\begin{equation}
	\left(-\frac{d^2}{d\rho^2}+\frac{\lambda_0(\rho)+15/4}{\rho^2}-Q_{00}(\rho)-\frac{2mE}{\hbar^2}\right)f_0(\rho)=0\;,
	\label{eq:5}
\end{equation}
with
\begin{equation}
	Q_{00}=\left\langle\Phi_0\left|\frac{\partial^2}{\partial\rho^2}\right|\Phi_0\right\rangle_\Omega\;,
	\label{eq:6}
\end{equation}
where brackets indicate integration over all hyperangles. 
Numerically $Q_{00}$ is found to be extremely small compared to other terms 
and we will not include it in further calculations. In the following we will 
suppress the subscript $0$. 

\subsection{Two-body Potential Models}
Our main concern is the introduction of finite range
effects beyond the single-parameter zero-range approximation (where only the 
scattering length is included) {\it but} still using only contact interactions. 
There are different ways to do this. Here we use
a boundary condition model (or range expansion model)
and a two-channel model, and compare to the usual zero-range
approximation with only the scattering length.
To make our discussion self-contained
and well-defined we now proceed to introduce first the scattering length only
zero-range model and then the two effective-range models.

\subsubsection{Zero-range Model.}
The so-called zero-range models generally use a contact interaction potential which is defined 
by a boundary condition on the logarithmic derivative of the wave function 
at zero separation. In hyperspherical coordinates this boundary condition 
becomes \cite{fj01b}
\begin{equation}
	\frac{\partial(\alpha_i\Phi)}{\partial\alpha_i}\bigg|_{\alpha_i=0}=-\sqrt2\rho\frac1{a_i}\alpha_i\Phi\bigg|_{\alpha_i=0}\;,
	\label{eq:7}
\end{equation}
where $a_i$ is the scattering length between particles $j$ and $k$. Since all particles are equal 
we will suppress indices on scattering variables such as $a$. 
Using Faddeev decomposition of the angular wave function with $s$-states only,
\begin{equation}
	\Phi=\phi_1+\phi_2+\phi_3\;,
	\label{eq:8}
\end{equation}
where 
\begin{equation}
	\phi_i=N(\rho)\sin\left(\nu\left[\alpha_i-\frac{\pi}{2}\right]\right)
	\label{eq:9}
\end{equation}
is a solution to \eqref{eq:4}, $\nu^2=\lambda+4$ and $N(\rho)$ is a normalization factor, we find the eigenvalue equation
\begin{equation}
	\frac{\nu\cos\left(\frac{\nu\pi}2\right)-\frac{8}{\sqrt3}\sin\left(\frac{\nu\pi}6\right)}{\sin\left(\frac{\nu\pi}2\right)}=\frac{\sqrt2\rho}{a}\;,
	\label{eq:10}
\end{equation}
after rotating two of the Faddeev components into the 
coordinate system of the third \cite{fj01b}. 
For large positive $\rho/a$ there is an asymptotic solution of the form
$\nu\rightarrow i\sqrt{2}\rho/a$ yielding a dimer binding energy of 
\begin{equation}
	E_D=\frac{\lambda+\frac{15}{4}}{2\rho^2}=\frac{\nu^2-\frac{1}{4}}{2\rho^2}\approx-\frac{1}{a^2}\;,
	\label{eq:11}
\end{equation}
(in units where $\hbar=m=1$). For negative $a$ there are no bound dimers. 
The limit $\rho/a=0$ yields the solution $\nu=is_0$ with $s_0=1.00624$. These are the
basic properties of the simplest single-parameter zero-range model when applied to a 
system of three identical bosons.

\subsubsection{Effective Range Expansion.}
The first method for including finite range effects beyond the scattering length
is the effective range expansion (range expansion or boundary condition 
model for short). 
It is a generalization of the boundary condition (\ref{eq:7}).
From scattering theory the effective range expansion 
of the phase shift $\delta$ as a function of wave-number $k$ is
\begin{equation}
	\lim_{k\rightarrow0}k\cot\delta(k) = -\frac{1}{a}+\frac12Rk^2\;,
	\label{eq:12}
\end{equation}
where $a$ is the scattering length and $R$ is known as the effective range. 
As it stands $R$ is an additional parameter. However, later we will take $R$ to dependent on the scattering length $a$, i.e. $R=R(a)$,
since this is the physical reality of Feshbach-resonances, where both $a$ and $R$ are dependent on the external magnetic field 
(this effect arises naturally in the two-channel model described below). When we use $R(a)$ in 
the effective range expansion we will be assuming the same functional dependence on $a$ as 
found in the two-channel model (see equation (\ref{eq:21})) since this is a generic
feature of many Feshbach resonance models \cite{chin2010}.

The boundary condition \eqref{eq:7} now becomes (see \cite{fj01b})
\begin{equation}
	\frac{\partial(\alpha_i\Phi)}{\partial\alpha_i}\bigg|_{\alpha_i=0}=
	\sqrt2\rho\left[-\frac1{a_i}+\frac12R_i \frac{\nu^2}{2\rho^2}\right]\alpha_i\Phi\bigg|_{\alpha_i=0}\;,
	\label{eq:13}
\end{equation}
where $R_i$ is the effective range between particles $j$ and $k$, and the momentum in \eqref{eq:12} is given by 
$k=\nu/(\sqrt2\rho)$ \cite{fj01b}. Again assuming that all particles are
equal, the eigenvalue equation \eqref{eq:10} becomes
\begin{equation}
\frac{\nu\cos\left(\frac{\nu\pi}2\right)-\frac{8}{\sqrt3}\sin\left(\frac{\nu\pi}6\right)}{\sin\left(\frac{\nu\pi}2\right)}=\sqrt2\rho\left[\frac{1}{a}-\frac{1}{2}R\left(\frac{\nu}{\sqrt2\rho}\right)^2\right].
	\label{eq:14}
\end{equation}
Inclusion of the effective range yields the dimer binding energy
\begin{equation}
	E_D=\frac{-1}{R^2}\left(1-\sqrt{1-2\frac{R}{a}}\right)^2\approx \frac{-1}{a^2}\left(1+\frac{R}{a}\right)\;,
	\label{eq:15}
\end{equation}
for $|R|\ll a$. Thus the dimer system can be more or less bound 
depending on the sign of the effective range. In the case of 
atomic Feshbach resonances the effective range is negative 
\cite{chin2010}, yielding less bound dimers.

This model breaks down for sufficiently large positive effective ranges
since there exist no solution to the eigenvalue equation \eqref{eq:14} for 
$\rho\lesssim R$ if $R$ is positive. To remedy this deficiency an additional parameter
known as the shape parameter, $P$, can be included in \eqref{eq:12} 
\cite{fj01b}
\begin{equation}
	\lim_{k\rightarrow0}k\cot\delta(k) = -\frac{1}{a}+\frac12Rk^2+PR^3k^4\;,
	\label{eq:16}
\end{equation}
and \eqref{eq:13} gets a similar additional term. Typical values of $P$ are around $0.1$.
However, we note that results from binding energy calculations are largely insensitive
to the exact value of $P$ \cite{fj01b}. We therefore fix $P=0.1$ in 
this study. We again stress that when using (\ref{eq:16}) we take $R=R(a)$ given in (\ref{eq:21}) in 
order to model Feshbach resonances.

\subsubsection{Two-channel Model.} The other model for including finite range effects that we consider
is a two-channel contact interaction model that takes the internal degrees of freedom of 
the atoms into account. This model has background scattering 
lengths in open 
and closed channels that we denote $a_\op$ and $a_\cl$ respectively. 
The full scattering length and effective range parameters of this model are (all details of this model are described in our previous work \cite{twochannelFeshbach})
\begin{align}
	\frac1a&=\frac1{a_\op}+\frac{\beta^2}{\kappa-\frac1{a_\cl}}\;,\label{eq:17}\\
	R&=\frac{-\beta^2}{\kappa\left(\kappa-\frac1{a_\cl}\right)^2}\;,
	\label{eq:18}
\end{align}
where $\beta$ parametrises the coup\-ling between the channels and $\kappa$ is given by the 
energy separation $E^*=\hbar^2\kappa^2/2m_r$ between the channels, where $m_r = \tfrac m2$ 
is the two-body reduced mass. The expressions are valid when $a\gg\kappa^{-1}$ 
(equivalently $E^*\gg\hbar^2/2m_ra^2$), which is always the case near a Feshbach resonance.
At the resonance, i.e. $a=\infty$, the effective range is \cite{chin2010}
\begin{equation}
	R_0 = -\frac1{a_{bg}}\frac{\hbar^2}{m_r\Delta\mu\Delta B} \;,
	\label{eq:19}
\end{equation}
where $\Delta B$ is the width of the resonance, $\Delta\mu$ the difference 
in magnetic moments of the channels and $a_{bg}$ the background scattering 
length (the value away from the given Feshbach resonance). 
The resonance width $\Delta B$ and position $B_0$ as given in the phenomenological expression \cite{chin2010}
\begin{equation}
	a(B) \approx a_{bg}\left(1 - \frac{\Delta B}{B-B_0} \right) \;,
	\label{eq:20}
\end{equation}
which can be derived in the two-channel model. The parameters $a_\op$, $a_\cl$, 
$\beta$ and $E^*$ can now be replaced by $\Delta B$, $B_0$, the difference in 
magnetic momenta $\Delta\mu$ and off-resonance scattering length $a_{bg}$ to relate
all quantities to physically measured values. Since we address neutral atom systems
here the long-range atom-atom interaction is of the van der Waals type and can 
be characterized by the van der Waals length, $r_\textrm{vdW}$. It has been 
shown that the scattering length in a potential with a van der Waals tail is 
$\sim 0.956r_\textrm{vdw}$ \cite{PhysRevA.48.1993}. For simplicity we will 
assume that $a_{bg}=r_\textrm{vdW}$ and use the two interchangeably. This also
means that we are working with $a_{bg}>0$ throughout this paper. We
have checked that the case with $a_{bg}<0$ gives the same qualitative results.
Off resonance the model predicts the effective range $R$ in terms of $R_0$ and $a$
\begin{equation}
	R(a) = R_0\left(1-\frac{a_{bg}}{a}\right)^2\;,
	\label{eq:21}
\end{equation}
similar to the results obtained in \cite{gao1998,PhysRevA.59.1998,zinner2009,thogersen2009,zinner2009b}.
Since the $a$-dependence
of $R$ of \eqref{eq:21} should be generic for Feshbach resonances (independent
of the particular scattering model used), we will use $R=R(a)$ in both the 
effective range expansion model \eqref{eq:12} and the two-channel model.

Note that the effective range is always negative since $R_0$ is always negative
for an atomic Feshbach resonance \cite{chin2010}, which in turn means that 
the two-channel model studied here will have $R(a)<0$ always. Multi-channel 
calculations of resonance properties have shown that some Feshbach resonances
can have positive $R$ in the vicinity of the resonance (see for instance 
reference \cite{gross2009} for an example in $^7$Li). This is typically 
only found for broad resonances (with $\Delta B$ large and correspondingly 
small $|R_0|$). For narrow resonances with $\Delta B$ small, 
$|R_0|$ is large and can potentially
cause non-negligible corrections to the zero-range models, this is the 
case we are interesting in here. Around the latter type of resonance 
the effective range, $R(a)$, is found to be negative. The two-channel model
we employ here is only applicable to those Feshbach resonances where the effective
range is negative. 

\subsection{Model Comparison}
\begin{figure}
\centering
% GNUPLOT: LaTeX picture with Postscript
\begingroup
  \makeatletter
  \providecommand\color[2][]{%
    \GenericError{(gnuplot) \space\space\space\@spaces}{%
      Package color not loaded in conjunction with
      terminal option `colourtext'%
    }{See the gnuplot documentation for explanation.%
    }{Either use 'blacktext' in gnuplot or load the package
      color.sty in LaTeX.}%
    \renewcommand\color[2][]{}%
  }%
  \providecommand\includegraphics[2][]{%
    \GenericError{(gnuplot) \space\space\space\@spaces}{%
      Package graphicx or graphics not loaded%
    }{See the gnuplot documentation for explanation.%
    }{The gnuplot epslatex terminal needs graphicx.sty or graphics.sty.}%
    \renewcommand\includegraphics[2][]{}%
  }%
  \providecommand\rotatebox[2]{#2}%
  \@ifundefined{ifGPcolor}{%
    \newif\ifGPcolor
    \GPcolorfalse
  }{}%
  \@ifundefined{ifGPblacktext}{%
    \newif\ifGPblacktext
    \GPblacktexttrue
  }{}%
  % define a \g@addto@macro without @ in the name:
  \let\gplgaddtomacro\g@addto@macro
  % define empty templates for all commands taking text:
  \gdef\gplbacktext{}%
  \gdef\gplfronttext{}%
  \makeatother
  \ifGPblacktext
    % no textcolor at all
    \def\colorrgb#1{}%
    \def\colorgray#1{}%
  \else
    % gray or color?
    \ifGPcolor
      \def\colorrgb#1{\color[rgb]{#1}}%
      \def\colorgray#1{\color[gray]{#1}}%
      \expandafter\def\csname LTw\endcsname{\color{white}}%
      \expandafter\def\csname LTb\endcsname{\color{black}}%
      \expandafter\def\csname LTa\endcsname{\color{black}}%
      \expandafter\def\csname LT0\endcsname{\color[rgb]{1,0,0}}%
      \expandafter\def\csname LT1\endcsname{\color[rgb]{0,1,0}}%
      \expandafter\def\csname LT2\endcsname{\color[rgb]{0,0,1}}%
      \expandafter\def\csname LT3\endcsname{\color[rgb]{1,0,1}}%
      \expandafter\def\csname LT4\endcsname{\color[rgb]{0,1,1}}%
      \expandafter\def\csname LT5\endcsname{\color[rgb]{1,1,0}}%
      \expandafter\def\csname LT6\endcsname{\color[rgb]{0,0,0}}%
      \expandafter\def\csname LT7\endcsname{\color[rgb]{1,0.3,0}}%
      \expandafter\def\csname LT8\endcsname{\color[rgb]{0.5,0.5,0.5}}%
    \else
      % gray
      \def\colorrgb#1{\color{black}}%
      \def\colorgray#1{\color[gray]{#1}}%
      \expandafter\def\csname LTw\endcsname{\color{white}}%
      \expandafter\def\csname LTb\endcsname{\color{black}}%
      \expandafter\def\csname LTa\endcsname{\color{black}}%
      \expandafter\def\csname LT0\endcsname{\color{black}}%
      \expandafter\def\csname LT1\endcsname{\color{black}}%
      \expandafter\def\csname LT2\endcsname{\color{black}}%
      \expandafter\def\csname LT3\endcsname{\color{black}}%
      \expandafter\def\csname LT4\endcsname{\color{black}}%
      \expandafter\def\csname LT5\endcsname{\color{black}}%
      \expandafter\def\csname LT6\endcsname{\color{black}}%
      \expandafter\def\csname LT7\endcsname{\color{black}}%
      \expandafter\def\csname LT8\endcsname{\color{black}}%
    \fi
  \fi
  \setlength{\unitlength}{0.0500bp}%
  \begin{picture}(7200.00,4536.00)%
    \gplgaddtomacro\gplbacktext{%
      \csname LTb\endcsname%
      \put(726,704){\makebox(0,0)[r]{\strut{}-2}}%
      \put(726,1596){\makebox(0,0)[r]{\strut{}-1.5}}%
      \put(726,2488){\makebox(0,0)[r]{\strut{}-1}}%
      \put(726,3379){\makebox(0,0)[r]{\strut{}-0.5}}%
      \put(726,4271){\makebox(0,0)[r]{\strut{} 0}}%
      \put(858,484){\makebox(0,0){\strut{} 0}}%
      \put(1651,484){\makebox(0,0){\strut{} 20}}%
      \put(2443,484){\makebox(0,0){\strut{} 40}}%
      \put(3236,484){\makebox(0,0){\strut{} 60}}%
      \put(4029,484){\makebox(0,0){\strut{} 80}}%
      \put(4821,484){\makebox(0,0){\strut{} 100}}%
      \put(5614,484){\makebox(0,0){\strut{} 120}}%
      \put(6407,484){\makebox(0,0){\strut{} 140}}%
      \put(3830,154){\makebox(0,0){\strut{}$\rho/a_\textrm{bg}$}}%
      \put(65,2488){\makebox(0,0)[l]{\strut{}$\nu_0^2$}}%
    }%
    \gplgaddtomacro\gplfronttext{%
      \csname LTb\endcsname%
      \put(990,1482){\makebox(0,0)[l]{\strut{}Zero-range}}%
      \csname LTb\endcsname%
      \put(990,1196){\makebox(0,0)[l]{\strut{}Two-channel}}%
      \csname LTb\endcsname%
      \put(990,910){\makebox(0,0)[l]{\strut{}Boundary condition}}%
    }%
    \gplgaddtomacro\gplbacktext{%
      \csname LTb\endcsname%
      \put(3834,2456){\makebox(0,0)[r]{\strut{}-10}}%
      \put(3834,2769){\makebox(0,0)[r]{\strut{}-8}}%
      \put(3834,3081){\makebox(0,0)[r]{\strut{}-6}}%
      \put(3834,3394){\makebox(0,0)[r]{\strut{}-4}}%
      \put(3834,3706){\makebox(0,0)[r]{\strut{}-2}}%
      \put(3834,4019){\makebox(0,0)[r]{\strut{} 0}}%
      \put(3966,2236){\makebox(0,0){\strut{} 0}}%
      \put(4461,2236){\makebox(0,0){\strut{} 200}}%
      \put(4957,2236){\makebox(0,0){\strut{} 400}}%
      \put(5452,2236){\makebox(0,0){\strut{} 600}}%
      \put(5948,2236){\makebox(0,0){\strut{} 800}}%
      \put(6443,2236){\makebox(0,0){\strut{} 1000}}%
      \put(3416,3237){\rotatebox{-270}{\makebox(0,0){\strut{}}}}%
      \put(5204,1950){\makebox(0,0){\strut{}}}%
    }%
    \gplgaddtomacro\gplfronttext{%
    }%
    \gplbacktext
    \put(0,0){\includegraphics{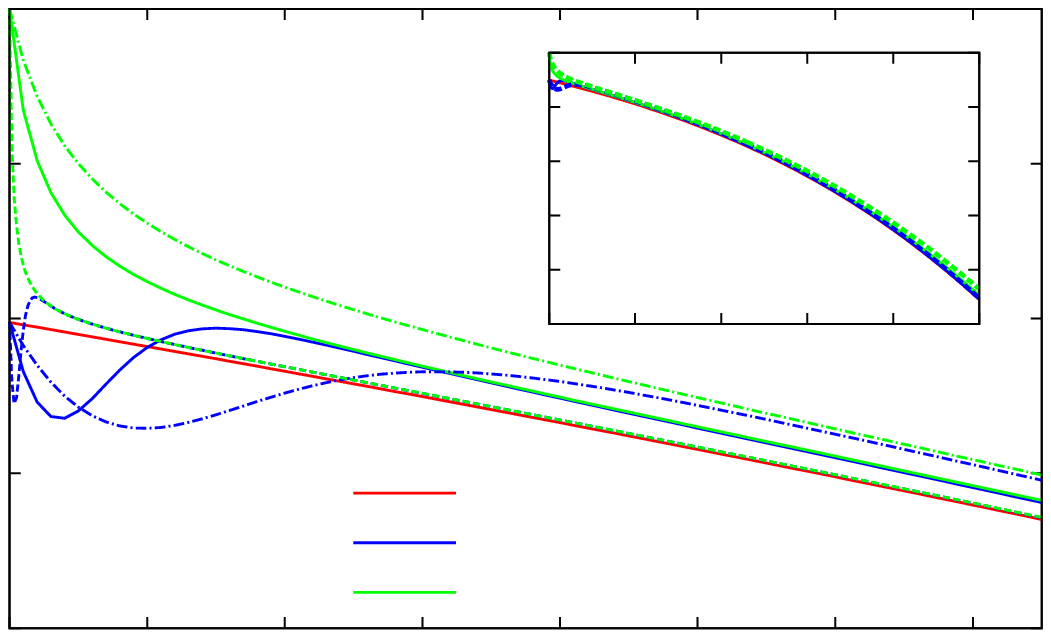}}%
    \gplfronttext
  \end{picture}%
\endgroup

\caption{Small and large $\rho$ behaviour of the hyperangular 
eigenvalues for $n=0$, for the three models with $a=500$, $R_0=-1$ 
(short dash), $R_0=-10$ (full lines) and $R_0=-25$ (dash dot), all in units of $a_{bg}$.}
\label{fig1}
\end{figure}

Before we consider the recombination rates and the binding energies in the 
different models, we first make a comparison of the two-body potential models
in terms of their predictions for the $\nu^2=\lambda+4$ coefficient that 
provides the effective potential for the three-body system in the hyperradial 
equation (\ref{eq:5}).
We compare the models by explicitly plotting their associated eigenvalues in 
figure~\ref{fig1}.
At large $\rho$ all models have the same asymptotic value 
$\nu^2=-2\rho^2/a^2$. This is because we have a two-body bound state for $a>0$, i.e.
the structure is that two particles are bound and have small interparticle distance while
the third particle is far away.
This is more clearly illustrated in the inset of figure~\ref{fig1}
where the horizontal axis extends up to $\rho/a_{bg}=1000$.
For intermediate distances $\rho\gtrsim2|R_0|$ the 
finite-range models show surprisingly similar forms given their 
quite different formalism. What is particularly important to notice is the 
fact that an inner pocket develops in the two-channel model, but at the 
same time a barrier with respect to the zero-range model is also seen. 
For $\rho<|R_0|$ the effective range expansion
model eigenvalue goes to zero, and thus effectively becomes regularized. There is no need for 
a three-body parameter (in the form of a high-momentum cut-off in momentum-space or 
a short-distance cut-off in coordinate space). This is not the case for 
the two-channel model we use here. It requires a cut-off at small distance when 
calculating e.g. bound state energies (\ref{eq:25}) and the recombination rate (\ref{eq:24}).
This can be understood by considering that the two-channel model consists of 
two single channel models that are coupled together by a coupling potential 
of zero-range. This means that there is no scale coming from the 
coupling that can regularize the three-body problem and in turn one still
needs to introduce a short-distance cut-off as discussed in reference \cite{twochannelFeshbach}.

\section{Finite-range effects in the recombination rate}\label{recRate}
We now proceed to consider three-body observables starting with the
recombination rate on the positive $a$ side of the Feshbach resonance.
On this side of the resonance the recombination takes place by transition
of two of the three particles into the channel with a bound two-body
dimer with the universal binding energy proportional to $-a^{-2}$. On the 
$a<0$ side where there is no bound dimer, the decay goes directly into 
some strongly bound two-body state of the atom-atom potential and depends
on the short-range details. This latter case will not be considered here.

The recombination rate for $a>0$ is calculated using the semi-classical 
WKB method 
of hidden crossing theory where the recombination coefficient $\alpha$ is \cite{Macek1999}
\begin{equation}
	\alpha=8(2\pi)^23\sqrt3\frac\hbar{m_r}\lim_{k\rightarrow0}\frac{P(k)}{k^4}\;,
	\label{eq:22}
\end{equation}
with the wave-number $k$ defined by $E=\hbar^2k^2/2m_r$. The transition probability $P(k)\equiv|S_{01}(k)|^2$, with $S_{01}$ the transition matrix element between adiabatic channels 0 and 1, is \cite{Macek1999}
\begin{align}
	P(k)&=4e^{-2\Sigma}\sin^2\Delta\label{eq:23}\;,\\
	\Delta+i\Sigma&=\int_{\mathcal{C}} d\rho\sqrt{k^2-\frac{\nu(\rho)^2}{\rho^2}}\;,
	\label{eq:24}
\end{align}
where the integral is taken along a contour $\mathcal{C}$ in the complex 
$\rho$-plane connecting the adiabatic channel corresponding to 
three free particles, $n=1$, to the channel describing a dimer 
and a free particle, $n=0$. The integration path $\mathcal{C}$ goes around a so-called branchpoint $\rho_b$ that connects the two channels. Additional details can be found in 
reference~\cite{twochannelFeshbach}.
Note that since $\nu_0^2(0)<0$ for the zero-range and two-channel 
models there is no classical turning point in the open ($n=1$) 
channel, thus we employ a regularization cut-off in order to 
avoid a divergent integral.

The perhaps more familiar form $\alpha = C(a)\hbar a^4/m$ (as for instance
found in reference \cite{braaten2006}) can be found from the above equations in 
the universal limit ($a=\infty$) of the single channel model (where 
$\nu(\rho)=is_0$). Here $C(a)$ is a log-periodic function of $a$.
We can now split the integral in two parts, one from the 
cutoff $\rho_\text{cut}$ to the real part of the branchpoint $\rho_\text{b}$ and 
another for the rest. If we denote the first part by $\Delta_1+i\Sigma_1$, then we 
have the result 
$\Delta_1 = s_0\log(\text{Re}(\rho_b)/\rho_\text{cut})$ and $\Sigma_1=0$. 
The vanishing of the imaginary part comes about since 
the potential is negative in the lower branch and $k^2$ is positive,
thus yielding a purely real integrand. Now, the branchpoint is 
simply related to $a$ by $\rho_b = (1.8327 + 2.1029i)a$ in the 
single channel model and thus $\text{Re}(\rho_b)\propto a$. When plugging 
this into (\ref{eq:23}) the log-periodic dependency is established. 
The rest of the integration path only leads to a constant phase-shift 
independent of $a$ since $\nu(\rho)$ only depends on the ratio $\rho/a$, 
see equation (\ref{eq:10}). The $a^4$ dependency is most easily seen 
by dimensional analysis, $k$ has units of inverse length and the 
only available length scale in the zero range model is the scattering length $a$.

\begin{figure}[h]
\centering
% GNUPLOT: LaTeX picture with Postscript
\begingroup
  \makeatletter
  \providecommand\color[2][]{%
    \GenericError{(gnuplot) \space\space\space\@spaces}{%
      Package color not loaded in conjunction with
      terminal option `colourtext'%
    }{See the gnuplot documentation for explanation.%
    }{Either use 'blacktext' in gnuplot or load the package
      color.sty in LaTeX.}%
    \renewcommand\color[2][]{}%
  }%
  \providecommand\includegraphics[2][]{%
    \GenericError{(gnuplot) \space\space\space\@spaces}{%
      Package graphicx or graphics not loaded%
    }{See the gnuplot documentation for explanation.%
    }{The gnuplot epslatex terminal needs graphicx.sty or graphics.sty.}%
    \renewcommand\includegraphics[2][]{}%
  }%
  \providecommand\rotatebox[2]{#2}%
  \@ifundefined{ifGPcolor}{%
    \newif\ifGPcolor
    \GPcolorfalse
  }{}%
  \@ifundefined{ifGPblacktext}{%
    \newif\ifGPblacktext
    \GPblacktexttrue
  }{}%
  % define a \g@addto@macro without @ in the name:
  \let\gplgaddtomacro\g@addto@macro
  % define empty templates for all commands taking text:
  \gdef\gplbacktext{}%
  \gdef\gplfronttext{}%
  \makeatother
  \ifGPblacktext
    % no textcolor at all
    \def\colorrgb#1{}%
    \def\colorgray#1{}%
  \else
    % gray or color?
    \ifGPcolor
      \def\colorrgb#1{\color[rgb]{#1}}%
      \def\colorgray#1{\color[gray]{#1}}%
      \expandafter\def\csname LTw\endcsname{\color{white}}%
      \expandafter\def\csname LTb\endcsname{\color{black}}%
      \expandafter\def\csname LTa\endcsname{\color{black}}%
      \expandafter\def\csname LT0\endcsname{\color[rgb]{1,0,0}}%
      \expandafter\def\csname LT1\endcsname{\color[rgb]{0,1,0}}%
      \expandafter\def\csname LT2\endcsname{\color[rgb]{0,0,1}}%
      \expandafter\def\csname LT3\endcsname{\color[rgb]{1,0,1}}%
      \expandafter\def\csname LT4\endcsname{\color[rgb]{0,1,1}}%
      \expandafter\def\csname LT5\endcsname{\color[rgb]{1,1,0}}%
      \expandafter\def\csname LT6\endcsname{\color[rgb]{0,0,0}}%
      \expandafter\def\csname LT7\endcsname{\color[rgb]{1,0.3,0}}%
      \expandafter\def\csname LT8\endcsname{\color[rgb]{0.5,0.5,0.5}}%
    \else
      % gray
      \def\colorrgb#1{\color{black}}%
      \def\colorgray#1{\color[gray]{#1}}%
      \expandafter\def\csname LTw\endcsname{\color{white}}%
      \expandafter\def\csname LTb\endcsname{\color{black}}%
      \expandafter\def\csname LTa\endcsname{\color{black}}%
      \expandafter\def\csname LT0\endcsname{\color{black}}%
      \expandafter\def\csname LT1\endcsname{\color{black}}%
      \expandafter\def\csname LT2\endcsname{\color{black}}%
      \expandafter\def\csname LT3\endcsname{\color{black}}%
      \expandafter\def\csname LT4\endcsname{\color{black}}%
      \expandafter\def\csname LT5\endcsname{\color{black}}%
      \expandafter\def\csname LT6\endcsname{\color{black}}%
      \expandafter\def\csname LT7\endcsname{\color{black}}%
      \expandafter\def\csname LT8\endcsname{\color{black}}%
    \fi
  \fi
  \setlength{\unitlength}{0.0500bp}%
  \begin{picture}(7200.00,4536.00)%
    \gplgaddtomacro\gplbacktext{%
      \csname LTb\endcsname%
      \put(924,704){\makebox(0,0)[r]{\strut{}$10^{-34}$}}%
      \put(924,1150){\makebox(0,0)[r]{\strut{}$10^{-32}$}}%
      \put(924,1596){\makebox(0,0)[r]{\strut{}$10^{-30}$}}%
      \put(924,2042){\makebox(0,0)[r]{\strut{}$10^{-28}$}}%
      \put(924,2488){\makebox(0,0)[r]{\strut{}$10^{-26}$}}%
      \put(924,2933){\makebox(0,0)[r]{\strut{}$10^{-24}$}}%
      \put(924,3379){\makebox(0,0)[r]{\strut{}$10^{-22}$}}%
      \put(924,3825){\makebox(0,0)[r]{\strut{}$10^{-20}$}}%
      \put(924,4271){\makebox(0,0)[r]{\strut{}$10^{-18}$}}%
      \put(1866,484){\makebox(0,0){\strut{}$10^2$}}%
      \put(4334,484){\makebox(0,0){\strut{}$10^3$}}%
      \put(6803,484){\makebox(0,0){\strut{}$10^4$}}%
      \put(22,2487){\rotatebox{-270}{\makebox(0,0){\strut{}$\alpha_\textrm{rec}/(\textrm{cm}^6\textrm{s}^{-1})$}}}%
      \put(3929,154){\makebox(0,0){\strut{}$a/a_\textrm{bg}$}}%
      \put(1968,2933){\makebox(0,0)[l]{\strut{}$a_1^*$}}%
      \put(5273,3825){\makebox(0,0)[l]{\strut{}$a_2^*$}}%
    }%
    \gplgaddtomacro\gplfronttext{%
      \csname LTb\endcsname%
      \put(1188,4065){\makebox(0,0)[l]{\strut{}Zero-range}}%
      \csname LTb\endcsname%
      \put(1188,3779){\makebox(0,0)[l]{\strut{}Two-channel}}%
      \csname LTb\endcsname%
      \put(1188,3493){\makebox(0,0)[l]{\strut{}Boundary condition}}%
    }%
    \gplgaddtomacro\gplbacktext{%
      \csname LTb\endcsname%
      \put(3804,1070){\makebox(0,0)[r]{\strut{}$10^{-33}$}}%
      \put(3804,1445){\makebox(0,0)[r]{\strut{}$10^{-31}$}}%
      \put(3804,1819){\makebox(0,0)[r]{\strut{}$10^{-29}$}}%
      \put(3804,2194){\makebox(0,0)[r]{\strut{}$10^{-27}$}}%
      \put(3936,850){\makebox(0,0){\strut{}80}}%
      \put(4807,850){\makebox(0,0){\strut{}110}}%
      \put(5656,850){\makebox(0,0){\strut{}150}}%
      \put(6443,850){\makebox(0,0){\strut{}200}}%
      \put(3122,1725){\rotatebox{-270}{\makebox(0,0){\strut{}}}}%
      \put(5189,564){\makebox(0,0){\strut{}}}%
    }%
    \gplgaddtomacro\gplfronttext{%
    }%
    \gplbacktext
    \put(0,0){\includegraphics{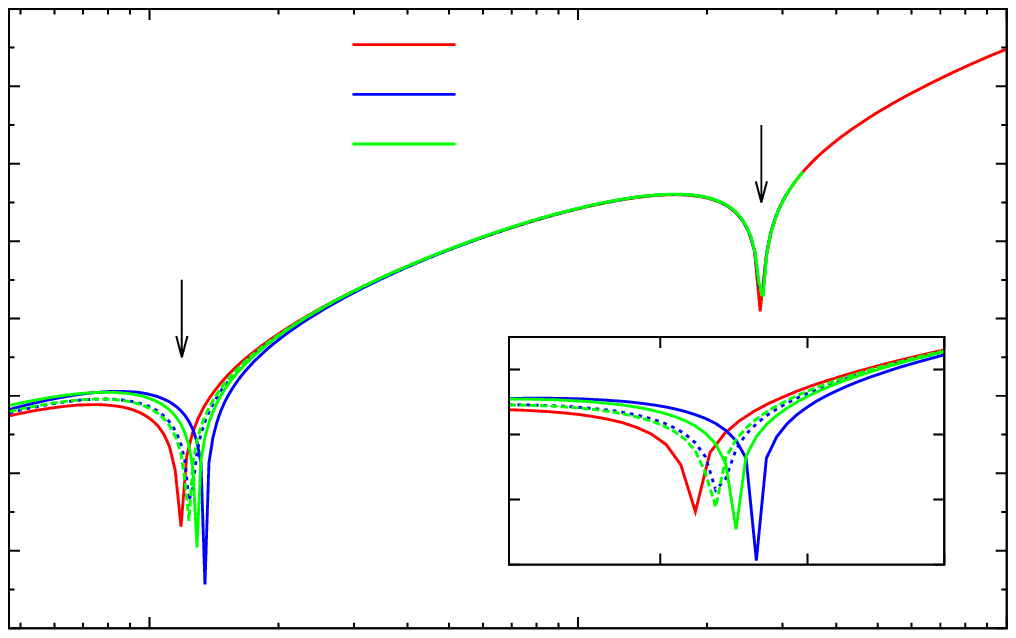}}%
    \gplfronttext
  \end{picture}%
\endgroup
\caption{Recombination coefficient $\alpha$ \eqref{eq:22} for the 
zero-range model, the effective range expansion model and 
the two-channel model with $R_0=-3a_{bg}$ 
(dotted) and $R_0=-10a_{bg}$ (full). The inset shows a closer look at 
the minimum near $a_1^*$. Note that the cut-off is such that
all models reproduce the minimum at $a_{2}^{*}$. This allows us to 
study the effect at the other minimum of the different models.}
\label{fig2}
\end{figure}

\begin{figure}[h]
\centering
% GNUPLOT: LaTeX picture with Postscript
\begingroup
  \makeatletter
  \providecommand\color[2][]{%
    \GenericError{(gnuplot) \space\space\space\@spaces}{%
      Package color not loaded in conjunction with
      terminal option `colourtext'%
    }{See the gnuplot documentation for explanation.%
    }{Either use 'blacktext' in gnuplot or load the package
      color.sty in LaTeX.}%
    \renewcommand\color[2][]{}%
  }%
  \providecommand\includegraphics[2][]{%
    \GenericError{(gnuplot) \space\space\space\@spaces}{%
      Package graphicx or graphics not loaded%
    }{See the gnuplot documentation for explanation.%
    }{The gnuplot epslatex terminal needs graphicx.sty or graphics.sty.}%
    \renewcommand\includegraphics[2][]{}%
  }%
  \providecommand\rotatebox[2]{#2}%
  \@ifundefined{ifGPcolor}{%
    \newif\ifGPcolor
    \GPcolorfalse
  }{}%
  \@ifundefined{ifGPblacktext}{%
    \newif\ifGPblacktext
    \GPblacktexttrue
  }{}%
  % define a \g@addto@macro without @ in the name:
  \let\gplgaddtomacro\g@addto@macro
  % define empty templates for all commands taking text:
  \gdef\gplbacktext{}%
  \gdef\gplfronttext{}%
  \makeatother
  \ifGPblacktext
    % no textcolor at all
    \def\colorrgb#1{}%
    \def\colorgray#1{}%
  \else
    % gray or color?
    \ifGPcolor
      \def\colorrgb#1{\color[rgb]{#1}}%
      \def\colorgray#1{\color[gray]{#1}}%
      \expandafter\def\csname LTw\endcsname{\color{white}}%
      \expandafter\def\csname LTb\endcsname{\color{black}}%
      \expandafter\def\csname LTa\endcsname{\color{black}}%
      \expandafter\def\csname LT0\endcsname{\color[rgb]{1,0,0}}%
      \expandafter\def\csname LT1\endcsname{\color[rgb]{0,1,0}}%
      \expandafter\def\csname LT2\endcsname{\color[rgb]{0,0,1}}%
      \expandafter\def\csname LT3\endcsname{\color[rgb]{1,0,1}}%
      \expandafter\def\csname LT4\endcsname{\color[rgb]{0,1,1}}%
      \expandafter\def\csname LT5\endcsname{\color[rgb]{1,1,0}}%
      \expandafter\def\csname LT6\endcsname{\color[rgb]{0,0,0}}%
      \expandafter\def\csname LT7\endcsname{\color[rgb]{1,0.3,0}}%
      \expandafter\def\csname LT8\endcsname{\color[rgb]{0.5,0.5,0.5}}%
    \else
      % gray
      \def\colorrgb#1{\color{black}}%
      \def\colorgray#1{\color[gray]{#1}}%
      \expandafter\def\csname LTw\endcsname{\color{white}}%
      \expandafter\def\csname LTb\endcsname{\color{black}}%
      \expandafter\def\csname LTa\endcsname{\color{black}}%
      \expandafter\def\csname LT0\endcsname{\color{black}}%
      \expandafter\def\csname LT1\endcsname{\color{black}}%
      \expandafter\def\csname LT2\endcsname{\color{black}}%
      \expandafter\def\csname LT3\endcsname{\color{black}}%
      \expandafter\def\csname LT4\endcsname{\color{black}}%
      \expandafter\def\csname LT5\endcsname{\color{black}}%
      \expandafter\def\csname LT6\endcsname{\color{black}}%
      \expandafter\def\csname LT7\endcsname{\color{black}}%
      \expandafter\def\csname LT8\endcsname{\color{black}}%
    \fi
  \fi
  \setlength{\unitlength}{0.0500bp}%
  \begin{picture}(7200.00,4032.00)%
    \gplgaddtomacro\gplbacktext{%
      \csname LTb\endcsname%
      \put(858,704){\makebox(0,0)[r]{\strut{} 18}}%
      \put(858,1010){\makebox(0,0)[r]{\strut{} 18.5}}%
      \put(858,1317){\makebox(0,0)[r]{\strut{} 19}}%
      \put(858,1623){\makebox(0,0)[r]{\strut{} 19.5}}%
      \put(858,1929){\makebox(0,0)[r]{\strut{} 20}}%
      \put(858,2236){\makebox(0,0)[r]{\strut{} 20.5}}%
      \put(858,2542){\makebox(0,0)[r]{\strut{} 21}}%
      \put(858,2848){\makebox(0,0)[r]{\strut{} 21.5}}%
      \put(858,3154){\makebox(0,0)[r]{\strut{} 22}}%
      \put(858,3461){\makebox(0,0)[r]{\strut{} 22.5}}%
      \put(858,3767){\makebox(0,0)[r]{\strut{} 23}}%
      \put(990,484){\makebox(0,0){\strut{}-12}}%
      \put(1959,484){\makebox(0,0){\strut{}-10}}%
      \put(2928,484){\makebox(0,0){\strut{}-8}}%
      \put(3897,484){\makebox(0,0){\strut{}-6}}%
      \put(4865,484){\makebox(0,0){\strut{}-4}}%
      \put(5834,484){\makebox(0,0){\strut{}-2}}%
      \put(6803,484){\makebox(0,0){\strut{} 0}}%
      \put(3896,154){\makebox(0,0){\strut{}$R_0/a_{bg}$}}%
      \put(-172,2236){\makebox(0,0)[l]{\strut{}$\dfrac{a_2^*}{a_1^*}$}}%
    }%
    \gplgaddtomacro\gplfronttext{%
      \csname LTb\endcsname%
      \put(4100,1482){\makebox(0,0)[l]{\strut{}Zero-range}}%
      \csname LTb\endcsname%
      \put(4100,1196){\makebox(0,0)[l]{\strut{}Two-channel}}%
      \csname LTb\endcsname%
      \put(4100,910){\makebox(0,0)[l]{\strut{}Boundary condition}}%
    }%
    \gplbacktext
    \put(0,0){\includegraphics{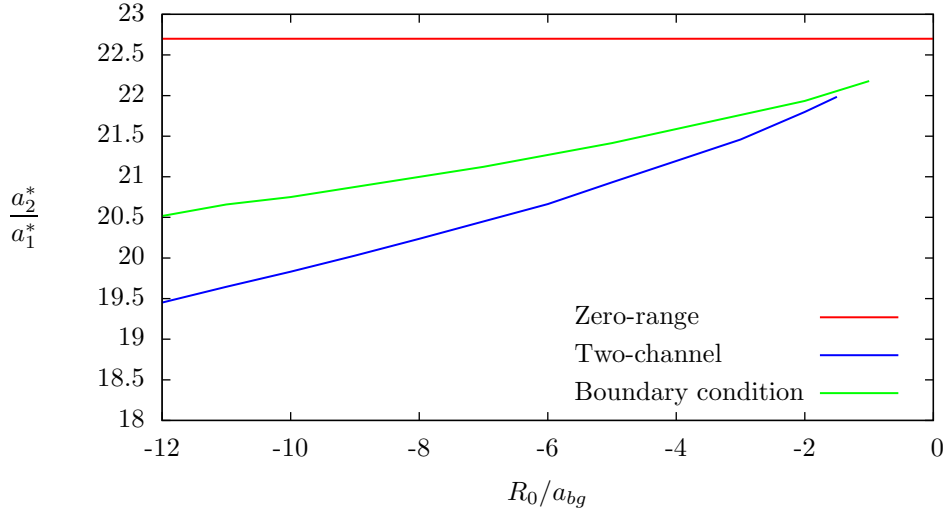}}%
    \gplfronttext
  \end{picture}%
\endgroup
\caption{The ratio $a_2^*/a_1^*$ for the zero-range model, the range expansion model
and the two-channel model as function of $R_0/a_{bg}$. The zero-range model has a ratio 
of 22.7. This factor is reduced when the range is increased to more negative values.}
\label{fig3}
\end{figure}

The recombination coefficients for different values of the effective 
ranges and different models are shown in figure~\ref{fig2}. The scattering 
length values $a_1^*$ and $a_2^*$ indicate locations of minima in the 
recombination rate. The minima are caused by the vanishing of bound 
trimers into the atom-dimer continuum.
The cut-offs were chosen such that 
the minimum at $a_2^*$ is the same for all models. We can thus compare
the models at the other minimum.
For the zero-range 
model, the ratio of $a_2^*$ to $a_1^*$ is $22.7$, showing that this 
calculation scheme agrees with the universal result. For the other 
models this ratio is reduced, the minimum at $a_1^*$ moves towards 
higher $a$. In order to make this more clear we plot the ratios
of the minima positions as function of the effective range, $R$, in
figure~\ref{fig3}. We see that the two-channel and range expansion
model give similar qualitative predictions but there are overall 
quantitative differences. We cannot extend the curves in figure~\ref{fig3}
all the way to $R_0=0$ due to numerical difficulties, but the trends
should be clear. What we also see is that the scale factor 
reduces quite drastically at large $R_0$ for both models. 
This corresponds to narrow Feshbach resonances, where
there are currently not enough experimental data to make a 
detailed comparison.

\subsection{Cut-off effects on binding energy}\label{cutoff}
As we noted already above, we need to put a cut-off on the WKB integral in order
to render it finite. We therefore need to consider the behaviour of the results as we change
this cut-off. In particular, the choice of cut-off affects the trimer binding energy. 
The WKB approximation yields a simple estimate of bound state energies (similar to 
the Bohr-Sommerfeld quantization rule)
\begin{equation}
	\int_{\rho_\text{cut}}^{\rho_\text{t}}d\rho\sqrt{2E_n-\frac{\nu_0(\rho)^2}{\rho^2}}=\pi\left(n-\frac{1}{4}\right)\;,
	\label{eq:25}
\end{equation}
where $n=1,2,\ldots$ indicate the ground state, first excited, $\ldots$ etc. $\rho_\text{cut}$ is defined at the end of Sec.~\ref{hyper}. It is required since the potential $\nu_0^2/\rho^2$ diverges as $\rho\rightarrow0$ and $\rho_\text{cut}$ acts as the innermost turningpoint in the WKB approximation. Likewise $\rho_t$ is the outermost turningpoint where $2E_n=\nu_0(\rho_t)^2/\rho_t^2$.

In the universal limit $a\rightarrow\infty$ where $\nu_0(\rho)=is_0$, $E_n$ is given by
\begin{equation}
	E_n\approx-\frac{2s_0^2}{\rho_\text{cut}^2} \exp\left(-\frac{2\pi n}{s_0}+\frac{\pi}{2s_0}-2\right)\;,
	\label{eq:26}
\end{equation}
clearly showing the geometric scaling $E_{n+1}=e^{-2\pi/s_0}E_n\approx E_n/22.7^2$ and the Thomas effect $E_n\rightarrow-\infty\text{ for }\rho_\text{cut}\rightarrow0$. 

Now we show that the cut-offs chosen such that the recombination minima at $a_2^*$ in figure~\ref{fig2} coincide for the three models lead to similar trimer binding energies. Solving \eqref{eq:25} numerically for finite scattering length gives trimer binding energies at e.g. $a=500, R=-10, n=2$ (in units of $\hbar^2/ma_\textrm{bg}^2$)
\begin{table}[h]
	\centering
	\begin{tabular}{lr}
		Zero-range: & $E_T$ = -0.002060\\
		Two-channel: & $E_T$ = -0.002441\\
		Effective range: & $E_T$ = -0.002062\\
	\end{tabular}
\end{table}

\noindent which shows that the chosen cut-offs give similar bound state energies for the different models. The 
effect of this choice of cut-off is thus under control.

\begin{figure}[h]
\centering
% GNUPLOT: LaTeX picture with Postscript
\begingroup
  \makeatletter
  \providecommand\color[2][]{%
    \GenericError{(gnuplot) \space\space\space\@spaces}{%
      Package color not loaded in conjunction with
      terminal option `colourtext'%
    }{See the gnuplot documentation for explanation.%
    }{Either use 'blacktext' in gnuplot or load the package
      color.sty in LaTeX.}%
    \renewcommand\color[2][]{}%
  }%
  \providecommand\includegraphics[2][]{%
    \GenericError{(gnuplot) \space\space\space\@spaces}{%
      Package graphicx or graphics not loaded%
    }{See the gnuplot documentation for explanation.%
    }{The gnuplot epslatex terminal needs graphicx.sty or graphics.sty.}%
    \renewcommand\includegraphics[2][]{}%
  }%
  \providecommand\rotatebox[2]{#2}%
  \@ifundefined{ifGPcolor}{%
    \newif\ifGPcolor
    \GPcolorfalse
  }{}%
  \@ifundefined{ifGPblacktext}{%
    \newif\ifGPblacktext
    \GPblacktexttrue
  }{}%
  % define a \g@addto@macro without @ in the name:
  \let\gplgaddtomacro\g@addto@macro
  % define empty templates for all commands taking text:
  \gdef\gplbacktext{}%
  \gdef\gplfronttext{}%
  \makeatother
  \ifGPblacktext
    % no textcolor at all
    \def\colorrgb#1{}%
    \def\colorgray#1{}%
  \else
    % gray or color?
    \ifGPcolor
      \def\colorrgb#1{\color[rgb]{#1}}%
      \def\colorgray#1{\color[gray]{#1}}%
      \expandafter\def\csname LTw\endcsname{\color{white}}%
      \expandafter\def\csname LTb\endcsname{\color{black}}%
      \expandafter\def\csname LTa\endcsname{\color{black}}%
      \expandafter\def\csname LT0\endcsname{\color[rgb]{1,0,0}}%
      \expandafter\def\csname LT1\endcsname{\color[rgb]{0,1,0}}%
      \expandafter\def\csname LT2\endcsname{\color[rgb]{0,0,1}}%
      \expandafter\def\csname LT3\endcsname{\color[rgb]{1,0,1}}%
      \expandafter\def\csname LT4\endcsname{\color[rgb]{0,1,1}}%
      \expandafter\def\csname LT5\endcsname{\color[rgb]{1,1,0}}%
      \expandafter\def\csname LT6\endcsname{\color[rgb]{0,0,0}}%
      \expandafter\def\csname LT7\endcsname{\color[rgb]{1,0.3,0}}%
      \expandafter\def\csname LT8\endcsname{\color[rgb]{0.5,0.5,0.5}}%
    \else
      % gray
      \def\colorrgb#1{\color{black}}%
      \def\colorgray#1{\color[gray]{#1}}%
      \expandafter\def\csname LTw\endcsname{\color{white}}%
      \expandafter\def\csname LTb\endcsname{\color{black}}%
      \expandafter\def\csname LTa\endcsname{\color{black}}%
      \expandafter\def\csname LT0\endcsname{\color{black}}%
      \expandafter\def\csname LT1\endcsname{\color{black}}%
      \expandafter\def\csname LT2\endcsname{\color{black}}%
      \expandafter\def\csname LT3\endcsname{\color{black}}%
      \expandafter\def\csname LT4\endcsname{\color{black}}%
      \expandafter\def\csname LT5\endcsname{\color{black}}%
      \expandafter\def\csname LT6\endcsname{\color{black}}%
      \expandafter\def\csname LT7\endcsname{\color{black}}%
      \expandafter\def\csname LT8\endcsname{\color{black}}%
    \fi
  \fi
  \setlength{\unitlength}{0.0500bp}%
  \begin{picture}(7200.00,4536.00)%
    \gplgaddtomacro\gplbacktext{%
      \csname LTb\endcsname%
      \put(946,704){\makebox(0,0)[r]{\strut{}-1}}%
      \put(946,1417){\makebox(0,0)[r]{\strut{}-0.8}}%
      \put(946,2131){\makebox(0,0)[r]{\strut{}-0.6}}%
      \put(946,2844){\makebox(0,0)[r]{\strut{}-0.4}}%
      \put(946,3558){\makebox(0,0)[r]{\strut{}-0.2}}%
      \put(946,4271){\makebox(0,0)[r]{\strut{} 0}}%
      \put(1078,484){\makebox(0,0){\strut{}-1}}%
      \put(2509,484){\makebox(0,0){\strut{}-0.5}}%
      \put(3941,484){\makebox(0,0){\strut{} 0}}%
      \put(5372,484){\makebox(0,0){\strut{} 0.5}}%
      \put(6803,484){\makebox(0,0){\strut{} 1}}%
      \put(176,2487){\rotatebox{-270}{\makebox(0,0){\strut{}$-\left\vert E_T\dfrac{ma_\textrm{bg}^{2}}{\hbar^2}\right\vert^{1/8}$}}}%
      \put(3940,154){\makebox(0,0){\strut{}$\textrm{sign}(a)\times\left\vert\dfrac{a_\textrm{bg}^2}{a^2}\right\vert^{1/8}$}}%
      \put(3654,1881){\makebox(0,0)[l]{\strut{}$n=1$}}%
      \put(3654,3272){\makebox(0,0)[l]{\strut{}$n=2$}}%
      \put(3941,3879){\makebox(0,0)[l]{\strut{}$3$}}%
      \put(1293,1524){\makebox(0,0)[l]{\strut{}$a^{(-)}$}}%
    }%
    \gplgaddtomacro\gplfronttext{%
      \csname LTb\endcsname%
      \put(2655,1482){\makebox(0,0)[l]{\strut{}Zero-range}}%
      \csname LTb\endcsname%
      \put(2655,1196){\makebox(0,0)[l]{\strut{}Two-channel}}%
      \csname LTb\endcsname%
      \put(2655,910){\makebox(0,0)[l]{\strut{}Boundary condition}}%
    }%
    \gplbacktext
    \put(0,0){\includegraphics{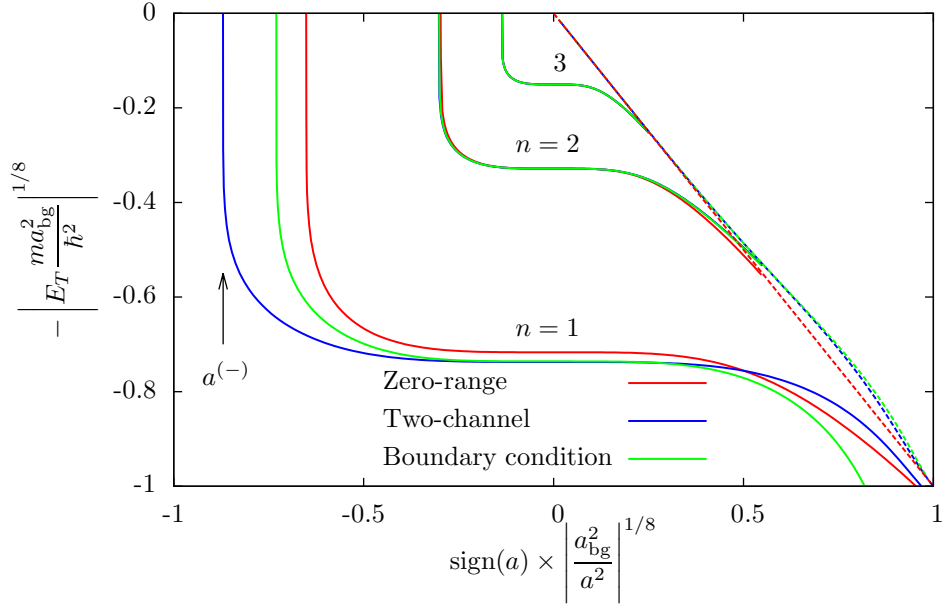}}%
    \gplfronttext
  \end{picture}%
\endgroup
\caption{Trimer bound state energy $E_T$ vs. inverse scattering length $a$ squared
for $R_0=-5r_\textrm{vdW}$. 
Both axes are scaled to the power $1/8$ to reasonably fit the entire spectrum on 
the plot. Dashed lines indicate atom-dimer threshold.}
\label{fig5}
\end{figure}

\section{Bound trimers}\label{boundTrimers}
In order to further study the three-body physics and its dependence
on two-body interaction models, we now consider the three-body bound
state spectrum when finite range effects are included.
When the scattering 
length $a$ is large, \eqref{eq:25} yields the same binding energy values 
as the radial equation \eqref{eq:5}. However, for the following discussion 
we will solve the radial equation numerically \eqref{eq:5} to obtain 
accurate results also when the binding energy is close to zero.
We solve the radial equation \eqref{eq:5} numerically for $f(\rho)$ with the boundary condition 
$f(\rho_\textrm{max}) = 0$ for some large hyperradius $\rho_\textrm{max}$.
The value of $\rho_\textrm{max}$ is chosen such that the bound state energy has 
converged to the desired degree of accuracy. 
Also, the aforementioned cut-off 
$\rho_\text{cut}$ is used in the boundary condition $f(\rho_\text{cut}) = 0$. 

For the three models in question we have calculated 
$E(a)$ for the three lowest trimers, the result is shown in figure~\ref{fig5}. 
For positive $a$ 
the dashed line indicates the atom-dimer thres\-hold which is given 
by the dimer binding energy $-1/a^2$ for the zero-range model and 
by \eqref{eq:15} for the effective range expansion model. For the 
two-channel model this can be calculated only numerically, yet it 
agrees surprisingly well with the analytical formula for the effective 
range expansion model \eqref{eq:15}.
The cut-offs (on the coordinate-space hyperspherical potential) 
were chosen such that the second trimer ($n=2$) energies 
coincide at $|a|=\infty$. The three spectra for $n=3$ are virtually 
identical. This is reasonable since the binding energy is very small for $n=3$ 
and the state is almost completely insensitive to finite-range effects. 
However, for the ground state $n=1$ a clear distinction between 
the models appears. At $|a|=\infty$ the finite-range models 
give practically the same trimer energy, a factor of $\sim25.3^2$ times 
higher than the $n=2$ state. In comparison the zero-range model trimer 
energy is only a factor of $22.7^2$ times higher.

We have already seen that the ratio of $a$-values corresponding 
to recombination minima in 
figure~\ref{fig2} differs from the universal value $22.7$. This is 
directly related to the change in the atom-dimer thres\-hold shown in 
figure~\ref{fig5} since the 
ratio between $a$ corresponding to trimer energy termination points on 
the threshold line is no longer $22.7$ for the models incorporating 
effective range effects. 

\begin{figure}[h]
\centering
% GNUPLOT: LaTeX picture with Postscript
\begingroup
  \makeatletter
  \providecommand\color[2][]{%
    \GenericError{(gnuplot) \space\space\space\@spaces}{%
      Package color not loaded in conjunction with
      terminal option `colourtext'%
    }{See the gnuplot documentation for explanation.%
    }{Either use 'blacktext' in gnuplot or load the package
      color.sty in LaTeX.}%
    \renewcommand\color[2][]{}%
  }%
  \providecommand\includegraphics[2][]{%
    \GenericError{(gnuplot) \space\space\space\@spaces}{%
      Package graphicx or graphics not loaded%
    }{See the gnuplot documentation for explanation.%
    }{The gnuplot epslatex terminal needs graphicx.sty or graphics.sty.}%
    \renewcommand\includegraphics[2][]{}%
  }%
  \providecommand\rotatebox[2]{#2}%
  \@ifundefined{ifGPcolor}{%
    \newif\ifGPcolor
    \GPcolorfalse
  }{}%
  \@ifundefined{ifGPblacktext}{%
    \newif\ifGPblacktext
    \GPblacktexttrue
  }{}%
  % define a \g@addto@macro without @ in the name:
  \let\gplgaddtomacro\g@addto@macro
  % define empty templates for all commands taking text:
  \gdef\gplbacktext{}%
  \gdef\gplfronttext{}%
  \makeatother
  \ifGPblacktext
    % no textcolor at all
    \def\colorrgb#1{}%
    \def\colorgray#1{}%
  \else
    % gray or color?
    \ifGPcolor
      \def\colorrgb#1{\color[rgb]{#1}}%
      \def\colorgray#1{\color[gray]{#1}}%
      \expandafter\def\csname LTw\endcsname{\color{white}}%
      \expandafter\def\csname LTb\endcsname{\color{black}}%
      \expandafter\def\csname LTa\endcsname{\color{black}}%
      \expandafter\def\csname LT0\endcsname{\color[rgb]{1,0,0}}%
      \expandafter\def\csname LT1\endcsname{\color[rgb]{0,1,0}}%
      \expandafter\def\csname LT2\endcsname{\color[rgb]{0,0,1}}%
      \expandafter\def\csname LT3\endcsname{\color[rgb]{1,0,1}}%
      \expandafter\def\csname LT4\endcsname{\color[rgb]{0,1,1}}%
      \expandafter\def\csname LT5\endcsname{\color[rgb]{1,1,0}}%
      \expandafter\def\csname LT6\endcsname{\color[rgb]{0,0,0}}%
      \expandafter\def\csname LT7\endcsname{\color[rgb]{1,0.3,0}}%
      \expandafter\def\csname LT8\endcsname{\color[rgb]{0.5,0.5,0.5}}%
    \else
      % gray
      \def\colorrgb#1{\color{black}}%
      \def\colorgray#1{\color[gray]{#1}}%
      \expandafter\def\csname LTw\endcsname{\color{white}}%
      \expandafter\def\csname LTb\endcsname{\color{black}}%
      \expandafter\def\csname LTa\endcsname{\color{black}}%
      \expandafter\def\csname LT0\endcsname{\color{black}}%
      \expandafter\def\csname LT1\endcsname{\color{black}}%
      \expandafter\def\csname LT2\endcsname{\color{black}}%
      \expandafter\def\csname LT3\endcsname{\color{black}}%
      \expandafter\def\csname LT4\endcsname{\color{black}}%
      \expandafter\def\csname LT5\endcsname{\color{black}}%
      \expandafter\def\csname LT6\endcsname{\color{black}}%
      \expandafter\def\csname LT7\endcsname{\color{black}}%
      \expandafter\def\csname LT8\endcsname{\color{black}}%
    \fi
  \fi
  \setlength{\unitlength}{0.0500bp}%
  \begin{picture}(7200.00,4032.00)%
    \gplgaddtomacro\gplbacktext{%
      \csname LTb\endcsname%
      \put(946,704){\makebox(0,0)[r]{\strut{}-1.9}}%
      \put(946,1044){\makebox(0,0)[r]{\strut{}-1.8}}%
      \put(946,1385){\makebox(0,0)[r]{\strut{}-1.7}}%
      \put(946,1725){\makebox(0,0)[r]{\strut{}-1.6}}%
      \put(946,2065){\makebox(0,0)[r]{\strut{}-1.5}}%
      \put(946,2406){\makebox(0,0)[r]{\strut{}-1.4}}%
      \put(946,2746){\makebox(0,0)[r]{\strut{}-1.3}}%
      \put(946,3086){\makebox(0,0)[r]{\strut{}-1.2}}%
      \put(946,3427){\makebox(0,0)[r]{\strut{}-1.1}}%
      \put(946,3767){\makebox(0,0)[r]{\strut{}-1}}%
      \put(1078,484){\makebox(0,0){\strut{}-2}}%
      \put(1794,484){\makebox(0,0){\strut{}-1.5}}%
      \put(2509,484){\makebox(0,0){\strut{}-1}}%
      \put(3225,484){\makebox(0,0){\strut{}-0.5}}%
      \put(3941,484){\makebox(0,0){\strut{} 0}}%
      \put(4656,484){\makebox(0,0){\strut{} 0.5}}%
      \put(5372,484){\makebox(0,0){\strut{} 1}}%
      \put(6087,484){\makebox(0,0){\strut{} 1.5}}%
      \put(6803,484){\makebox(0,0){\strut{} 2}}%
      \put(176,2235){\rotatebox{-270}{\makebox(0,0){\strut{}$a^{(-)}\kappa^*$}}}%
      \put(3940,154){\makebox(0,0){\strut{}$R_0/a_{bg}$}}%
    }%
    \gplgaddtomacro\gplfronttext{%
      \csname LTb\endcsname%
      \put(4100,3561){\makebox(0,0)[l]{\strut{}Zero-range}}%
      \csname LTb\endcsname%
      \put(4100,3275){\makebox(0,0)[l]{\strut{}Two-channel}}%
      \csname LTb\endcsname%
      \put(4100,2989){\makebox(0,0)[l]{\strut{}Boundary condition}}%
    }%
    \gplbacktext
    \put(0,0){\includegraphics{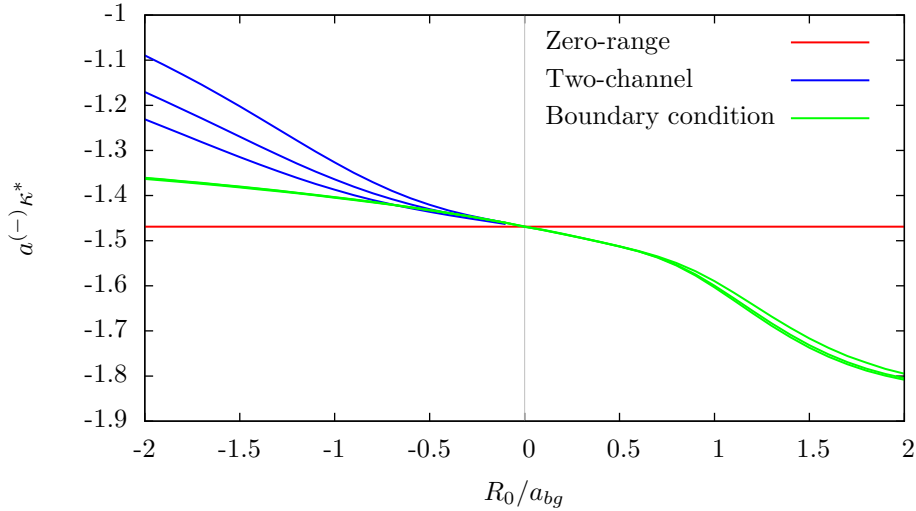}}%
    \gplfronttext
  \end{picture}%
\endgroup
\caption{The product $\kappa^*a^{(-)}$ as a function of effective 
range, $R_0$, where $-\hbar^2\kappa^{*2}/2m_r=E_T^{(\infty)}$, 
$E_T^{(\infty)}$ is the trimer binding energy at resonance ($a=\infty$) 
and $a^{(-)}$ is the threshold scattering length for trimer creation (see figure~\ref{fig5}). 
The universal value $\kappa^*a^{(-)}=-1.5076$ for the 
zero-range model is not correct for the lowest trimer, where the 
value is $-1.469$, independent of cutoff. The two-channel model 
curves are for different value coordinate-space cut-offs on the 
hyperradial potential. The cut-off is $0.5,\,0.6$ and $0.7$ in 
units of $a_{bg}$ for the top, middle and bottom blue curves. For 
the effective range expansion model the dependency on cutoff is 
insignificant. Note that the two-channel model only works for 
$R_0<0$.}
\label{fig6}
\end{figure}

For negative $a$ the value of $a^{(-)}$ indicates the thres\-hold 
scattering length for creation of the lowest Efimov trimer, as 
indicated in figure~\ref{fig5}. When written in units of $r_\textrm{vdW}$, 
this quantity is the subject of much recent discussion since it seems
to have a universal value of $a^{(-)}\sim -9r_\textrm{vdW}$ for 
different cold atomic systems \cite{berninger2011}. Here we want to 
address finite range effects on the value of $a^{(-)}$ for the  
two models. Some other recent works that address
finite range effects on this threshold value can be found in 
reference \cite{tfj08c} and \cite{naidon2011}.

Our results within the different models for $a^{(-)}$ as a function of 
$R_0/a_{bg}$ are shown in figure~\ref{fig6}. Most notable 
is the lowering of $a^{(-)}$ for the finite-range models 
compared to the zero-range model. This is partly due to 
the lower binding energy $E_T^{(\infty)}$ at $|a|=\infty$. The product 
$a^{(-)}\kappa^*$ (where $-\hbar^2\kappa^{*2}/2m_r=E_T^{(\infty)}$) is 
universal in the zero-range model with the value 
-1.5076 \cite{PhysRevLett.100.140404}. Thus increasing $|E_T^{(\infty)}|$ 
will reduce $|a^{(-)}|$. This effect is, however, not enough to account for 
the deviation from the zero-range result. The product 
is further reduced for decreasing $R_0$ indicating a lower value of $a^{(-)}$.

Both finite-range models show the same trend. 
However, the value of the change is different for the two models when $-R_0$ 
gets sufficiently large. The plot also shows that for the two-channel model 
the effect depends on cut-off. The cut-off, chosen for the purpose of 
illustration but with reasonable values, is $0.5,\,0.6$ and $0.7$ for the top,
middle and bottom blue curves in figure~\ref{fig6}. The 
effective range expansion model shows only a very small dependence on the cut-off 
(the three green curves in figure~\ref{fig6} show very little deviation).
The opposite behaviour, i.e. $|a^{(-)}\kappa^*|$ gets larger for larger effective 
range, is found for $R_0>0$. 
To obtain results for this case 
we use the boundary condition including the shape parameter 
\eqref{eq:16} in the effective range expansion model in order to extend it
to positive effective range. Indeed we find that the change is now 
opposite, see figure \ref{fig6}. This is similar to the product
of $a^{(-)}\kappa^*$ found in reference~\cite{schmidt2012}. We notice, 
however, that reference~\cite{schmidt2012} find that 
the trimer binding energies for $|a|=\infty$ get smaller and the 
value of $|a^{(-)}|$ gets larger for large effective range. This is 
not seen in our calculations. 
This can be connected to 
the use of finite-range potentials with positive effective range. 

\begin{figure}[ht!]
\centering
% GNUPLOT: LaTeX picture with Postscript
\begingroup
  \makeatletter
  \providecommand\color[2][]{%
    \GenericError{(gnuplot) \space\space\space\@spaces}{%
      Package color not loaded in conjunction with
      terminal option `colourtext'%
    }{See the gnuplot documentation for explanation.%
    }{Either use 'blacktext' in gnuplot or load the package
      color.sty in LaTeX.}%
    \renewcommand\color[2][]{}%
  }%
  \providecommand\includegraphics[2][]{%
    \GenericError{(gnuplot) \space\space\space\@spaces}{%
      Package graphicx or graphics not loaded%
    }{See the gnuplot documentation for explanation.%
    }{The gnuplot epslatex terminal needs graphicx.sty or graphics.sty.}%
    \renewcommand\includegraphics[2][]{}%
  }%
  \providecommand\rotatebox[2]{#2}%
  \@ifundefined{ifGPcolor}{%
    \newif\ifGPcolor
    \GPcolorfalse
  }{}%
  \@ifundefined{ifGPblacktext}{%
    \newif\ifGPblacktext
    \GPblacktexttrue
  }{}%
  % define a \g@addto@macro without @ in the name:
  \let\gplgaddtomacro\g@addto@macro
  % define empty templates for all commands taking text:
  \gdef\gplbacktext{}%
  \gdef\gplfronttext{}%
  \makeatother
  \ifGPblacktext
    % no textcolor at all
    \def\colorrgb#1{}%
    \def\colorgray#1{}%
  \else
    % gray or color?
    \ifGPcolor
      \def\colorrgb#1{\color[rgb]{#1}}%
      \def\colorgray#1{\color[gray]{#1}}%
      \expandafter\def\csname LTw\endcsname{\color{white}}%
      \expandafter\def\csname LTb\endcsname{\color{black}}%
      \expandafter\def\csname LTa\endcsname{\color{black}}%
      \expandafter\def\csname LT0\endcsname{\color[rgb]{1,0,0}}%
      \expandafter\def\csname LT1\endcsname{\color[rgb]{0,1,0}}%
      \expandafter\def\csname LT2\endcsname{\color[rgb]{0,0,1}}%
      \expandafter\def\csname LT3\endcsname{\color[rgb]{1,0,1}}%
      \expandafter\def\csname LT4\endcsname{\color[rgb]{0,1,1}}%
      \expandafter\def\csname LT5\endcsname{\color[rgb]{1,1,0}}%
      \expandafter\def\csname LT6\endcsname{\color[rgb]{0,0,0}}%
      \expandafter\def\csname LT7\endcsname{\color[rgb]{1,0.3,0}}%
      \expandafter\def\csname LT8\endcsname{\color[rgb]{0.5,0.5,0.5}}%
    \else
      % gray
      \def\colorrgb#1{\color{black}}%
      \def\colorgray#1{\color[gray]{#1}}%
      \expandafter\def\csname LTw\endcsname{\color{white}}%
      \expandafter\def\csname LTb\endcsname{\color{black}}%
      \expandafter\def\csname LTa\endcsname{\color{black}}%
      \expandafter\def\csname LT0\endcsname{\color{black}}%
      \expandafter\def\csname LT1\endcsname{\color{black}}%
      \expandafter\def\csname LT2\endcsname{\color{black}}%
      \expandafter\def\csname LT3\endcsname{\color{black}}%
      \expandafter\def\csname LT4\endcsname{\color{black}}%
      \expandafter\def\csname LT5\endcsname{\color{black}}%
      \expandafter\def\csname LT6\endcsname{\color{black}}%
      \expandafter\def\csname LT7\endcsname{\color{black}}%
      \expandafter\def\csname LT8\endcsname{\color{black}}%
    \fi
  \fi
  \setlength{\unitlength}{0.0500bp}%
  \begin{picture}(7200.00,4032.00)%
    \gplgaddtomacro\gplbacktext{%
      \csname LTb\endcsname%
      \put(814,704){\makebox(0,0)[r]{\strut{} 21}}%
      \put(814,1215){\makebox(0,0)[r]{\strut{} 22}}%
      \put(814,1725){\makebox(0,0)[r]{\strut{} 23}}%
      \put(814,2236){\makebox(0,0)[r]{\strut{} 24}}%
      \put(814,2746){\makebox(0,0)[r]{\strut{} 25}}%
      \put(814,3257){\makebox(0,0)[r]{\strut{} 26}}%
      \put(814,3767){\makebox(0,0)[r]{\strut{} 27}}%
      \put(946,484){\makebox(0,0){\strut{}-5}}%
      \put(1678,484){\makebox(0,0){\strut{}-4}}%
      \put(2410,484){\makebox(0,0){\strut{}-3}}%
      \put(3142,484){\makebox(0,0){\strut{}-2}}%
      \put(3875,484){\makebox(0,0){\strut{}-1}}%
      \put(4607,484){\makebox(0,0){\strut{} 0}}%
      \put(5339,484){\makebox(0,0){\strut{} 1}}%
      \put(6071,484){\makebox(0,0){\strut{} 2}}%
      \put(6803,484){\makebox(0,0){\strut{} 3}}%
      \put(176,2235){\rotatebox{-270}{\makebox(0,0){\strut{}$\sqrt{E^{(n)}/E^{(n+1)}}$}}}%
      \put(3874,154){\makebox(0,0){\strut{}$R_0/a_\textrm{bg}$}}%
    }%
    \gplgaddtomacro\gplfronttext{%
      \csname LTb\endcsname%
      \put(5024,3561){\makebox(0,0)[l]{\strut{}$r_c=0.1$}}%
      \csname LTb\endcsname%
      \put(5024,3275){\makebox(0,0)[l]{\strut{}$r_c=0.07$}}%
      \csname LTb\endcsname%
      \put(5024,2989){\makebox(0,0)[l]{\strut{}$r_c=0.05$}}%
      \csname LTb\endcsname%
      \put(5024,2703){\makebox(0,0)[l]{\strut{}$r_c=0.1$}}%
      \csname LTb\endcsname%
      \put(5024,2417){\makebox(0,0)[l]{\strut{}$r_c=0.07$}}%
      \csname LTb\endcsname%
      \put(5024,2131){\makebox(0,0)[l]{\strut{}$r_c=0.05$}}%
    }%
    \gplbacktext
    \put(0,0){\includegraphics{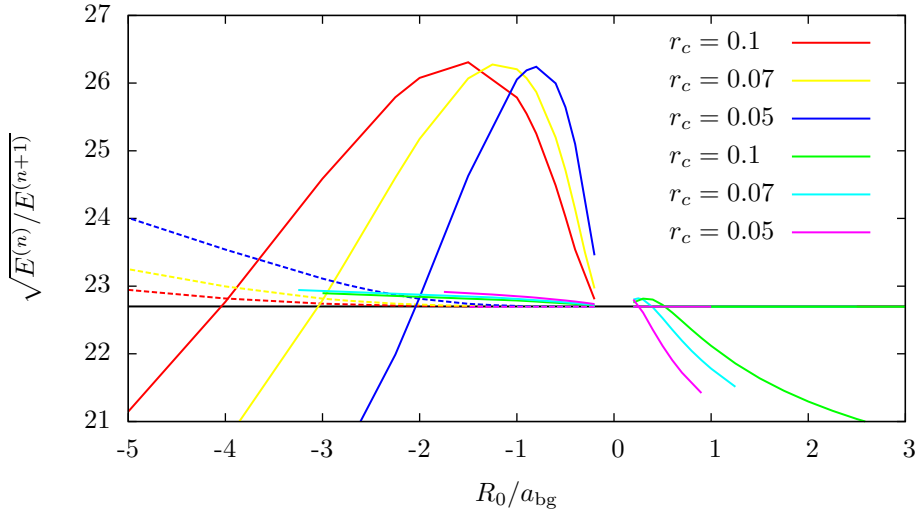}}%
    \gplfronttext
  \end{picture}%
\endgroup
\caption{Ratio of trimer bound state energies at resonance as a function of 
effective range for several different cut-offs. The solid lines show the 
ratio of energies for the first and second state, the dashed lines for the 
second and third state. Red, yellow and blue curves (top three in legend) 
are for the two channel model while green, cyan and magenta curves 
(bottom three in legend) are for the range expansion model.}
\label{fig7}
\end{figure}

In order to better understand the behaviour of the trimer energies and 
$a^{(-)}$, figure~\ref{fig7} shows the ratio of trimer bound state 
energies on resonance ($|a|\to\infty$) for 
the two-channel and range expansion models for three different cut-offs, chosen 
for the purpose of illustration. The most noticeable feature is the rise and fall 
of the ratio of the ground state energy $E^{(1)}$ to the first excited state 
energy $E^{(2)}$ for the two-channel model for negative $R_0$.
This non-monotonous behaviour can be understood if one assumes a 
three-body wave function that lives at large hyperradii, $\rho$.
When the effective range is decreased from zero 
($|R|$ increases) a barrier with respect to the pure zero-range model 
initially decreases
the binding energy. This can be clearly seen in figure~\ref{fig1}. 
As we increase the effective range further the wave function will leak 
into the attractive pocket at small $\rho$, which will again increase the 
binding compared to the pure zero-range result. This effect is 
strong for the ratio of the two lowest trimers but becomes weak for the 
ratio of the two highest trimers. This is understandable since the least
bound trimer resides at very large hyperradii and is largely insensitive
to the changes in the hyperradial potential at small $\rho$ due to 
the effective range correction.

\begin{figure}[h]
\centering
% GNUPLOT: LaTeX picture with Postscript
\begingroup
  \makeatletter
  \providecommand\color[2][]{%
    \GenericError{(gnuplot) \space\space\space\@spaces}{%
      Package color not loaded in conjunction with
      terminal option `colourtext'%
    }{See the gnuplot documentation for explanation.%
    }{Either use 'blacktext' in gnuplot or load the package
      color.sty in LaTeX.}%
    \renewcommand\color[2][]{}%
  }%
  \providecommand\includegraphics[2][]{%
    \GenericError{(gnuplot) \space\space\space\@spaces}{%
      Package graphicx or graphics not loaded%
    }{See the gnuplot documentation for explanation.%
    }{The gnuplot epslatex terminal needs graphicx.sty or graphics.sty.}%
    \renewcommand\includegraphics[2][]{}%
  }%
  \providecommand\rotatebox[2]{#2}%
  \@ifundefined{ifGPcolor}{%
    \newif\ifGPcolor
    \GPcolorfalse
  }{}%
  \@ifundefined{ifGPblacktext}{%
    \newif\ifGPblacktext
    \GPblacktexttrue
  }{}%
  % define a \g@addto@macro without @ in the name:
  \let\gplgaddtomacro\g@addto@macro
  % define empty templates for all commands taking text:
  \gdef\gplbacktext{}%
  \gdef\gplfronttext{}%
  \makeatother
  \ifGPblacktext
    % no textcolor at all
    \def\colorrgb#1{}%
    \def\colorgray#1{}%
  \else
    % gray or color?
    \ifGPcolor
      \def\colorrgb#1{\color[rgb]{#1}}%
      \def\colorgray#1{\color[gray]{#1}}%
      \expandafter\def\csname LTw\endcsname{\color{white}}%
      \expandafter\def\csname LTb\endcsname{\color{black}}%
      \expandafter\def\csname LTa\endcsname{\color{black}}%
      \expandafter\def\csname LT0\endcsname{\color[rgb]{1,0,0}}%
      \expandafter\def\csname LT1\endcsname{\color[rgb]{0,1,0}}%
      \expandafter\def\csname LT2\endcsname{\color[rgb]{0,0,1}}%
      \expandafter\def\csname LT3\endcsname{\color[rgb]{1,0,1}}%
      \expandafter\def\csname LT4\endcsname{\color[rgb]{0,1,1}}%
      \expandafter\def\csname LT5\endcsname{\color[rgb]{1,1,0}}%
      \expandafter\def\csname LT6\endcsname{\color[rgb]{0,0,0}}%
      \expandafter\def\csname LT7\endcsname{\color[rgb]{1,0.3,0}}%
      \expandafter\def\csname LT8\endcsname{\color[rgb]{0.5,0.5,0.5}}%
    \else
      % gray
      \def\colorrgb#1{\color{black}}%
      \def\colorgray#1{\color[gray]{#1}}%
      \expandafter\def\csname LTw\endcsname{\color{white}}%
      \expandafter\def\csname LTb\endcsname{\color{black}}%
      \expandafter\def\csname LTa\endcsname{\color{black}}%
      \expandafter\def\csname LT0\endcsname{\color{black}}%
      \expandafter\def\csname LT1\endcsname{\color{black}}%
      \expandafter\def\csname LT2\endcsname{\color{black}}%
      \expandafter\def\csname LT3\endcsname{\color{black}}%
      \expandafter\def\csname LT4\endcsname{\color{black}}%
      \expandafter\def\csname LT5\endcsname{\color{black}}%
      \expandafter\def\csname LT6\endcsname{\color{black}}%
      \expandafter\def\csname LT7\endcsname{\color{black}}%
      \expandafter\def\csname LT8\endcsname{\color{black}}%
    \fi
  \fi
  \setlength{\unitlength}{0.0500bp}%
  \begin{picture}(7200.00,4032.00)%
    \gplgaddtomacro\gplbacktext{%
      \csname LTb\endcsname%
      \put(726,704){\makebox(0,0)[r]{\strut{}-5.4}}%
      \put(726,1044){\makebox(0,0)[r]{\strut{}-5.3}}%
      \put(726,1385){\makebox(0,0)[r]{\strut{}-5.2}}%
      \put(726,1725){\makebox(0,0)[r]{\strut{}-5.1}}%
      \put(726,2065){\makebox(0,0)[r]{\strut{}-5}}%
      \put(726,2406){\makebox(0,0)[r]{\strut{}-4.9}}%
      \put(726,2746){\makebox(0,0)[r]{\strut{}-4.8}}%
      \put(726,3086){\makebox(0,0)[r]{\strut{}-4.7}}%
      \put(726,3427){\makebox(0,0)[r]{\strut{}-4.6}}%
      \put(726,3767){\makebox(0,0)[r]{\strut{}-4.5}}%
      \put(858,484){\makebox(0,0){\strut{} 0}}%
      \put(1849,484){\makebox(0,0){\strut{} 5}}%
      \put(2840,484){\makebox(0,0){\strut{} 10}}%
      \put(3831,484){\makebox(0,0){\strut{} 15}}%
      \put(4821,484){\makebox(0,0){\strut{} 20}}%
      \put(5812,484){\makebox(0,0){\strut{} 25}}%
      \put(6803,484){\makebox(0,0){\strut{} 30}}%
      \put(3830,154){\makebox(0,0){\strut{}$\rho/a_\textrm{bg}$}}%
      \put(-131,2065){\makebox(0,0)[l]{\strut{}$\lambda(\rho)$}}%
    }%
    \gplgaddtomacro\gplfronttext{%
      \csname LTb\endcsname%
      \put(3308,3594){\makebox(0,0)[l]{\strut{}One-channel}}%
      \csname LTb\endcsname%
      \put(3308,3374){\makebox(0,0)[l]{\strut{}Two-channel, $R_0<0$}}%
      \csname LTb\endcsname%
      \put(3308,3154){\makebox(0,0)[l]{\strut{}Boundary condition, $R_0<0$}}%
      \csname LTb\endcsname%
      \put(3308,2934){\makebox(0,0)[l]{\strut{}Boundary condition, $R_0>0$}}%
    }%
    \gplbacktext
    \put(0,0){\includegraphics{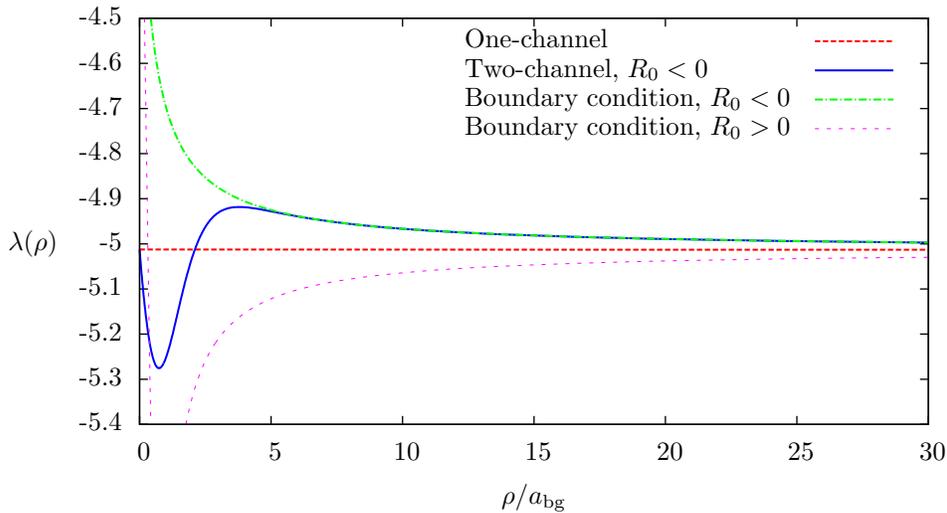}}%
    \gplfronttext
  \end{picture}%
\endgroup
\caption{The eigenvalue solutions to \eqref{eq:4} as a function of 
hyperradius, $\rho$ for the zero-range , two-channel model and range 
expansion model for a representative case with $|R_0|=5a_{bg}$. 
The lowest adiabatic potentials multiplied by $\rho^2$
as functions of hyperradius, $\rho$, for the zero range one-channel model
and three different effective ranges for the two-channel model.}
\label{fig4}
\end{figure}

With this interpretation the behaviour 
can now be better understood. In figure~\ref{fig4}
we plot the solution to \eqref{eq:4} which determines the 
effective three-body hyperradial potential for the different models
and for different signs of the effective range. The finite 
range corrections for the two-channel model and the effective 
range models are very different
as one has the inner pocket and the other does not. This is 
the origin of the differences seen in figure~\ref{fig7}. Initially
we see a repulsive effect compared to the pure zero-range model
that drives the ratio upwards as all states are pushed out to large
distance. However, for the two-channel model we eventually feel 
the presence of the inner pocket and states will leak into smaller
hyperradii where the ratio of energies is in turn driven down. 
For the positive effective range case we see from figure~\ref{fig4}
that the range expansion model will now have a pocket at small distance.
The trimer states can leak into this pocket and the ratio of energies
will again go down. The important point is that in the models we have
presented here there is no reason to expect monotonic behaviour 
since the effective three-body hyperradial potentials have 
non-trivial structure with repulsive and attractive parts in comparison
to the pure zero-range model with no effective range corrections.

\section{Conclusion}\label{Conclusion}
We investigate finite-range effects in three-body recombination rates in 
cold atomic gases near Feshbach resonances as well as finite-range effects 
in the trimer bound state energy spectrum. We use two models which include 
the finite-range effects and compare their results with a contact 
interaction (i.e. zero-range) model.
The first model is the range expansion model which is a straightforward 
extension of the zero-range contact interaction model. Here the effective 
range is included directly in the boundary condition on the 
three-body wave function following 
the effective range expansion of standard scattering theory. Variation of the 
scattering length through the Feshbach resonance is done phenomenologically 
as in the zero-range model. This model can also be used for positive 
effective range calculations. The second model is a two-channel contact 
interaction model which naturally 
includes both the finite effective range and the variation of the scattering 
length through the Feshbach resonance.

We show that with these well-tested two-body interaction models the three-body
physics can display complicated non-monotonic behaviour as the effective 
range is varied. In particular, we find that the geometric scaling 
factor of 22.7 for equal mass particles changes when including range
corrections, and that it can become both larger and smaller than 22.7 
depending on the magnitude and sign of the effective range.

In the current setup this can be understood based on the functional form of the 
effective hyperradial potential. On resonance where the 
scattering length diverges, the lowest trimer bound state has the strongest 
dependency on effective range since it lives at small hyperradius, whereas the 
excited states live at much larger hyperradii and the effective range contribution 
is much less pronounced. The adiabatic potential of the range 
expansion model is raised and lowered relative to the zero-range model potentials 
when the effective range is negative and positive, respectively. This leads to 
bound states being less or more bound, respectively. For the two-channel model 
the effective range is always negative which can only be achieved by using
potentials with an outer barrier. The hyperradial potential reflects this 
fact and develops a pocket at small hyperradii that the lowest states will eventually
leak into. This feature is similar to the range expansion model {\it but} for
the case of positive effective range. 

Our results demonstrate that effective range corrections within the framework of contact model potentials 
can lead to non-trivial behaviour of trimer energies, thresholds and interference features in 
recombination rates. Effective range corrections are expected to be important for the 
case of narrow Feshbach resonances. The experimental data on
Efimov states for narrow resonance systems is sparse and more measurements are 
needed in order to fully discriminate between different models that include 
finite range corrections. However, what we can conclude is that care must be 
taken when a particular two-body scattering model is used for the 
trimer states that have the largest binding energies in a universal 
setup, i.e. for the lowest states that have binding energies related to 
the background short-range scales. For higher lying trimers it is less
important since the states are largely insensitive to the short-distance
behaviour of the effective three-body potential.

\paragraph*{Acknowledgements} This work was supported by the Danish 
Agency for Science, Technology, and Innovation under the 
Danish Council for Independent Research - Natural Sciences.

\end{document}